\documentclass[useAMS,usenatbib]{mn2e}
\usepackage{hyperref}
\usepackage{graphicx}
\usepackage{subfig}

\title[Observational constraints on Emergent Universe]{Observational Constraints on the Model Parameters of a Class of Emergent Universe}
\author[S. Ghose, B. C. Paul and P. Thakur]{S. Ghose$^{1}$ \thanks{E-mail:souviknbu@rediffmail.com;}, P. Thakur $^{2}$ \thanks{E-mail: prasenjit_thakur1@yahoo.co.in} and B. C. Paul $^{1}$ \thanks{E-mail: bcpaul@iucaa.ernet.in} \\
$^{1}$ Department of Physics\\
University of North Bengal\\
Dist: Darjeelingm PIN - 734013\\
West Bengal, India\\
$^{2}$ Alipurduar College\\
Alipurduar, Dist: Jalpaiguri, PIN - 736122\\
West Bengal, India}

\begin{document}



\maketitle

\label{firstpage}

\begin{abstract}
A class of Emergent Universe (EU) model is studied in the light of recent observational data. Significant constraints on model parameters are obtained from the observational data. Density parameter for a class of model is evaluated.  Some of the models are in favour of the recent observations. Some models have been found which are not interesting yielding unrealistic present day value of the density parameter. 
\end{abstract}


\section{Introduction}

It is generally believed today that we live in an expanding Universe. After the discovery of CMBR \citep{penin, dicke} the big-bang cosmology has become the standard model for cosmology which accommodates a beginning of the Universe at some finite past. However, on its own big-bang cosmology does face some problems both in early and late universe. A number of problems cripped up when one describes the early universe, namely, the horizon problem, flatness problem etc. The above problems can be resolved evoking a phase of inflation \citep{guth,sato,linde,stein} at a very early epoch. On the other hand recent observations predict that our universe is passing through a phase of acceleration \citep{riess}. This phase of acceleration is believed to be a late time phase of the universe and may be accommodated in the standard model with a positive cosmological constant . Despite its overwhelming success, modern big-bang cosmology still has some unresolved issues. The physics of the inflation and introduction of a small cosmological constant for late acceleration, is not clearly understood \citep{reval,sean}. This is why there is enough motivation to search for alternative cosmology. Emergent universe (EU) models are employed to find a model which would accommodate the early inflationary phase and avoid the messy situation of the initial singularity \citep{ellis,har}. EU scenario can be realized in the framework of general relativity \citep{eu}, Gauss-Bonnet gravity \citep{eugb}, Brane world gravity \citep{b1,deb}, Brans-Dicke theory \citep{eubd} etc. Emergent universe are late time de-Sitter and thus naturally incorporate the late time accelerating phase as well. One such model was proposed by \citet{eu} in which a polytropic equation of state (EOS)in the form :
\begin{equation}
\label{eos1}
p= A \rho - B \rho^{1/2}.
\end{equation}
where $A$ and $B$ are constants is used. This is a special case of a more general equation
\begin{equation}
\label{eos2}
p= A\rho -B \rho^{\alpha}
\end{equation}
with $\alpha=1/2$. For such EOS a phenomenological construction can be found in string theories where most of the time models interpolate between two phases of universe \citep{fabris}. Universe in this model can stay large enough to avoid quantum gravitational effects even in the very early universe. Recently \citet{eu1} studied the viability of this type of model in the light of recent observational data and established bounds on model parameters $A$ and $B$. It is shown that the best fit value for $A$ may be very small but negative although a small positive value is allowed with 95$\%$ confidence. For a viable cosmology the bounds on $A$ and $B$ are determined for some fixed value of $K$. The parameter $K$, however, appears in the theory as an integration constant and may be fixed to some other value for different initial configuration. \citet{eu2} recently  worked with a more specific model for a small value of $A$ ($A \approx 0$). In the original work of \citet{eu} it was shown that the choice of $A$ drastically changes the matter energy composition of the Universe that eq. (\ref{eos1}) can mimic. In the present paper we obtain observational bounds on the model parameters $B$ and $K$ for different choices of $A$ as was considered in \citep{eu}. These choices correspond to very different compositions of cosmic fluid and it would be interesting to see whether realistic cosmologies are permitted for each case since the theory itself puts some constraints over $B$ and $K$ \citep{eu}].
The paper is presented as follows : in the next section we describe the relevant field equations. In section three we discuss the methods applied to constrain the parameters from (i) Observed Hubble Data (OHD)\citep{stern} and (ii) SDSS data measuring a model independent BAO peak parameter \citep{bao}. In section four we study the Density Parameters (DP) of the model (at the present epoch) and finally we discuss the results in section 5.
\section[]{Field Equations}
\begin{figure}
\centering
\subfloat[Part 1][]{\includegraphics[width=230pt,height=190pt]{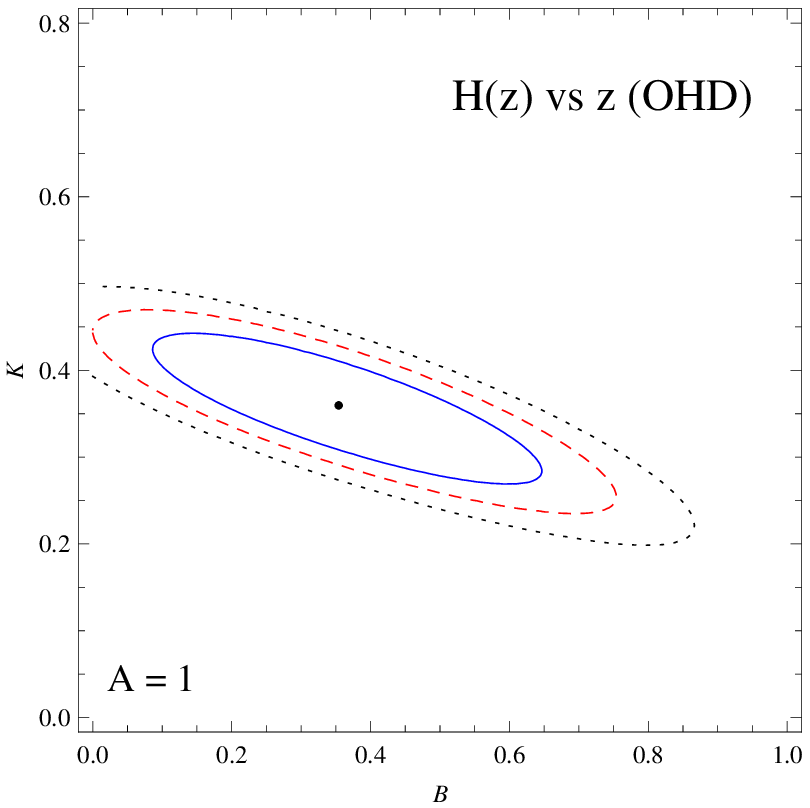} \label{ohd-a}}\\
\subfloat[Part 2][]{\includegraphics[width=230pt,height=190pt]{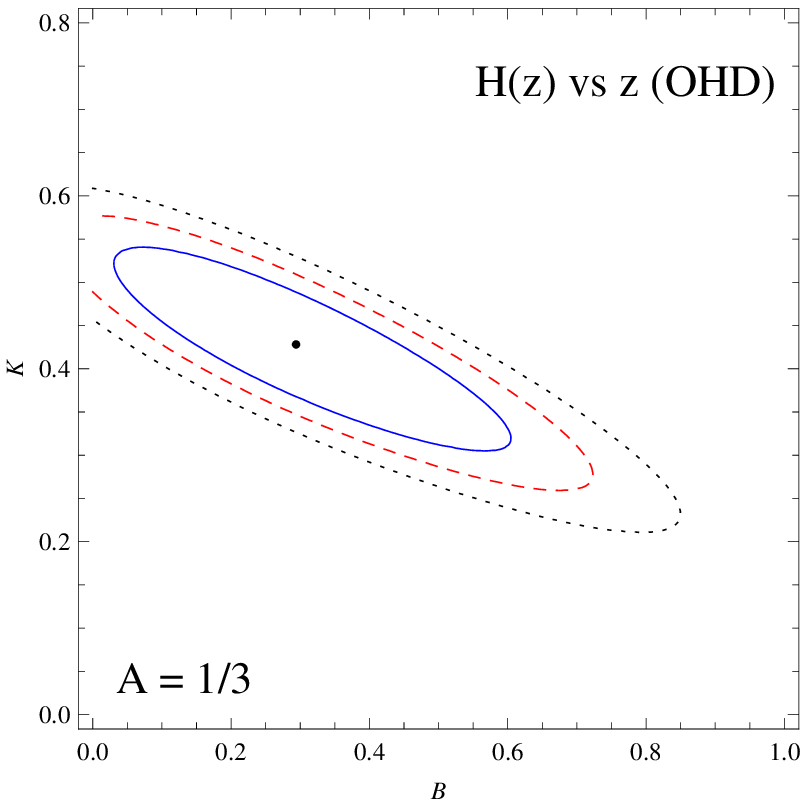} \label{ohd-b}}\\
\subfloat[Part 3][]{\includegraphics[width=230pt,height=190pt]{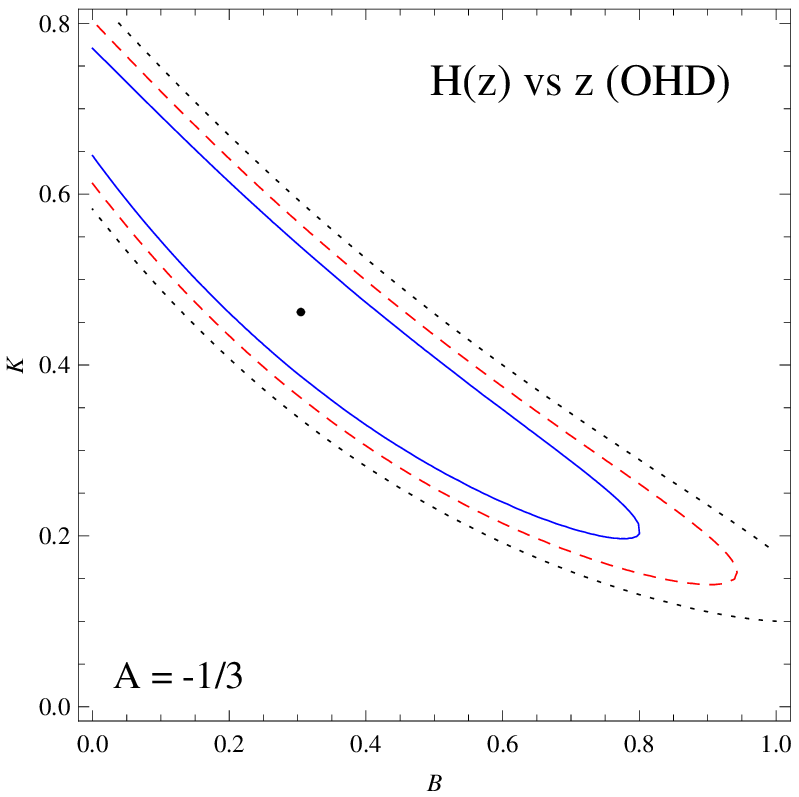} \label{ohd-c}}\\
\caption[Constraints from Stern Data]{(Colour Online)Constraints from OHD \citep{stern} Data for (a) $A=1$ (b) $A=1/3$ (c) $A=-1/3$:  $66.7\%$(Solid) $95.5\%$ (Dashed) and $99.8 \%$  (Dotted) contours.}
\label{ohd}
\end{figure}
Friedmann equation in a flat universe reads as:
\begin{equation}
\label{fr1}
H^{2}=\left(\frac{\dot{a}}{a}\right)^{2}=\frac{8 \pi G \rho}{3}
\end{equation}
where $H$ is the Hubble parameter and $a$ is the scale factor of the Universe.The usual conservation equation holds:
\begin{equation}
\label{csv}
\frac{d\rho}{dt}+3 H \left( p+\rho \right)
\end{equation} 
Using the EOS given by eq. (\ref{eos1}) in eq.(\ref{fr1}) and eq. (\ref{csv}) one obtains:
\begin{eqnarray}
\label{rho1}
\rho \left(z\right)=\left(\frac{B}{A+1}\right)^{2} +\frac{2 B K}{\left(A+1\right)^2} \left(1+z\right)^{\frac{3\left(A+1\right)}{2}} \\ \nonumber +\left(\frac{K}{A+1}\right)^{2}\left(1+z\right)^{3 \left(A+1\right)} 
\end{eqnarray}
where '$z$' represents the cosmological redshift. The first term in the right hand side of eq.(\ref{rho1}) is a constant which can be interpreted as cosmological constant and describing dark energy. Eq. (\ref{rho1}) can be written as:
\begin{equation}
\label{rho2}
\rho \left(z\right)=\rho_{1} +\rho_{2} \left(1+z\right)^{\frac{3\left(A+1\right)}{2}}+\rho_{3} \left(1+z\right)^{3 \left(A+1\right)}
\end{equation}
where $\rho_{1}=\left(\frac{B}{A+1}\right)^{2}$, $\rho_{2}=\frac{2 B K}{\left(A+1\right)^2}$ and $\rho_{3}=\left(\frac{K}{A+1}\right)^{2}$ represents densities at the present epoch. The Friedmann equation (eq. \ref{fr1}) can now be written in terms of redshift and density parameter as follows:
\begin{equation}
\label{fr2}
H^{2}\left(z\right)= H_{0}^{2}\left(\Omega_{1} +\Omega_{2} \left(1+z\right)^{\frac{3\left(A+1\right)}{2}}+\Omega_{3} \left(1+z\right)^{3 \left(A+1\right)}\right)
\end{equation}
where we define density parameter: $\Omega=\frac{8 \pi G \rho}{3H_{0}^{2}}=\Omega\left(A, B, K\right)$. For a given $A=A_{0}$ (say) we note that the nature of evolution for the variable parts of the matter energy density may now be established. Hence, choice of a suitable value for $A$ leads to a known composition of fluids. For example, \citet{eu2} considered the case $A=0$  with dark energy, dark matter and dust in the Universe. Fixing $A$ one can re-write eq. (\ref{fr2}) as:
\begin{equation}
\label{fr3}
H^{2}\left(H_{0}, B, K, z \right)= H_{0}^{2}E^{2}\left(B, K, z \right)
\end{equation}
where,
\begin{equation}
\label{fre}
E^{2}\left(B, K, z \right)=\Omega_{\Lambda} +\Omega_{2} \left(1+z\right)^{\frac{3\left(A+1\right)}{2}}+\Omega_{3} \left(1+z\right)^{3 \left(A+1\right)}.
\end{equation}
Here we have replaced the constant part of the DP ($\Omega_{1}$) by a new notation $\Omega_{\Lambda}$.
\section{Analysis with observational data}\subsection{Observed Hubble Data (OHD)} 
Using observed value of Hubble parameter at different redshifts (twelve data points listed in Observed Hubble Data by \citet{stern}) we analyse the model in this section. For the analysis we first define a chi square function as follows:
\begin{equation}
\label{chih1}
\chi^{2}_{OHD}=\sum \frac{\left(H_{Theory}\left(H_{0}, B, K, z\right)-H_{Obs}\right)^{2}}{2\sigma^{2}}
\end{equation}
where $H_{Theory}$ and $H_{Obs}$ are theoretical and observational values of Hubble parameter at different redshifts respectively and $\sigma$ is the corresponding error. Here, $H_{0}$ is a nuisance parameter and can be safely marginalized. We consider $H_{0}=72 \pm 8$ and a fixed prior distribution. A reduced chi square function can be defined as follows:
\begin{equation}
\label{chih2}
\chi^{2}_{red}=-2 ln \int \left[e^{-\frac{\chi^{2}_{OHD}}{2}}P\left(H_{0}\right)\right]dH_{0}
\end{equation}
where $P\left(H_{0}\right)$ is the prior distribution. The graph is plotted with $66.7\%$ (solid), $95.5\%$ (dashed) and $99.8\%$ (dotted) confidence level. The corresponding contours are shown in fig. (\ref{ohd}) for different $A$ values. The best fit values are tabulated in table (\ref{thub}).
\begin{table}
\centering
		\begin{tabular}{@{}lccc}
		\hline
		Model & $B$ & $K$ & $\chi^2_{min}$ (d.o.f)\\
		\hline
		$A=1$ & 0.354 & 0.360 &1.036 \\
		$A=1/3$ & 0.284 & 0.428 &1.030 \\
		$A=-1/3$ & 0.305 & 0.462 &1.019 \\
		\hline	
		\end{tabular}
\caption{Findings: OHD}	
	\label{thub}
\end{table}
\subsection{Joint analysis with BAO peak parameter}
\begin{figure}
\centering
\subfloat[Part 1][]{\includegraphics[width=230pt,height=190pt]{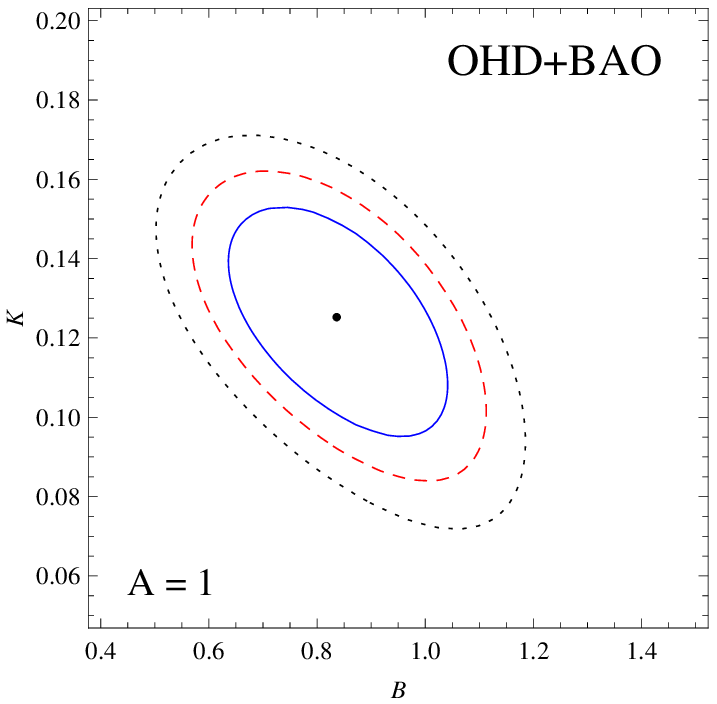} \label{bao-a}}\\
\subfloat[Part 2][]{\includegraphics[width=230pt,height=190pt]{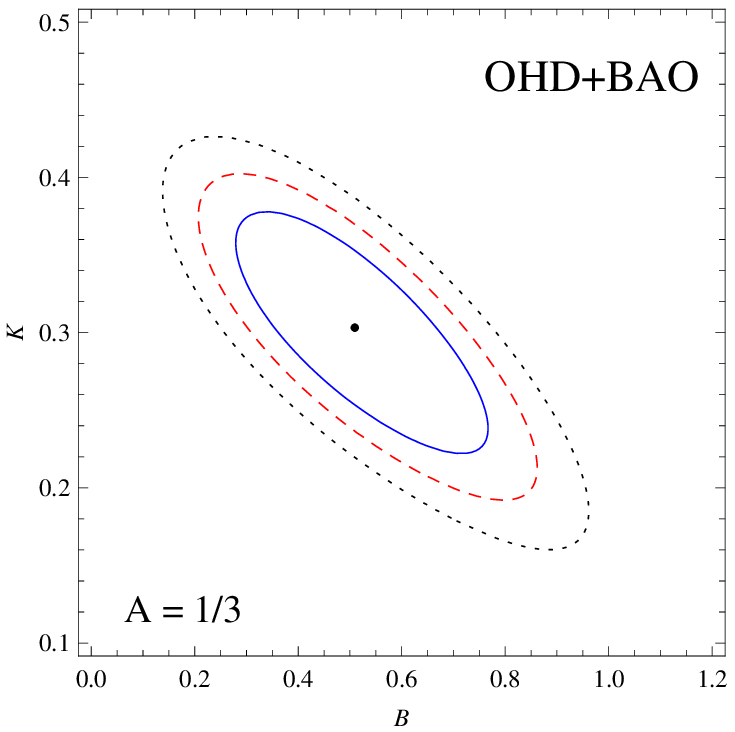} \label{bao-b}}\\
\subfloat[Part 3][]{\includegraphics[width=230pt,height=190pt]{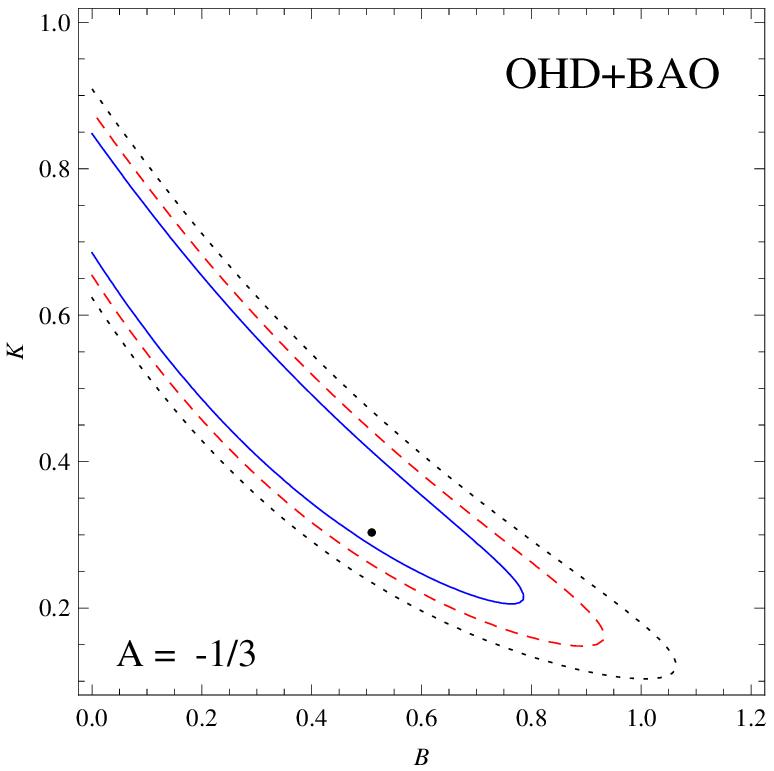} \label{bao-c}}\\
\caption[Constraints from OHD+SDSS Data]{(Colour Online)Constraints from joint OHD and SDSS(BAO) Data for (a) $A=1$ (b) $A=1/3$ (c) $A=-1/3$:  $66.7\%$(Solid) $95.5\%$ (Dashed) and $99.8 \%$  (Dotted) contours.}
\label{bao}
\end{figure}
In the previous analysis we used the standard value for $H_{0}$.In this section we consider analysis which is independent of the measurement of $H_{0}$ and does not consider any particular dark energy model. We use here a  method proposed by \citet{bao} and for this part of our analysis we follow their approach. A model independent BAO (Baryon Acuostic Oscillation) peak parameter can be defined for low redshift ($z_{1}$) measurements in a flat universe:
\begin{equation}
\label{baop}
\mathcal{A}=\frac{\Omega_{m}}{E\left(z_{1}\right)}\frac{\int_{0}^{z_{1}}\frac{dz}{E\left(z\right)}}{z_{1}}
\end{equation}
where $\Omega_{m}$ is the matter density parameter for the Universe. Now the chi square function can be defined as follows:
\begin{equation}
\label{chib}
\chi^{2}_{BAO}=\frac{\left(\mathcal{A}-0.469\right)^{2}}{2 \left(0.017\right)^{2}}
\end{equation}
where we have used the measured value for $\mathcal{A}$ ($0.469\pm.0.017$) as was obtained by \citet{bao} from the SDSS data for LRG (Luminous Red Galaxies) survey. Now we can define a total chi square function for our joint analysis as:
\begin{equation}
\label{chit}
\chi^{2}_{tot}=\chi^{2}_{red}+\chi^{2}_{BAO}
\end{equation}
The $66.7\%$ (solid), $95.5\%$ (dashed) and $99.8\%$ (dotted) contours obtained from this joint analysis is given in fig.(\ref{bao}). Best fit values are shown in table (\ref{tbao}).
\begin{table}
\centering
		\begin{tabular}{@{}lccc}
		\hline
		Model & $B$ & $K$ & $\chi^2_{min}$ (d.o.f)\\
		\hline
		$A=1$ & 0.836 & 0.125 & 1.888\\
		$A=1/3$ & 0.509 & 0.303 & 1.303 \\
		$A=-1/3$ & 0.210 & 0.560 & 1.401\\
		\hline	
		\end{tabular}
	\caption{Findings: OHD+SDSS (BAO)}
	\label{tbao}
\end{table}
\section{Analysis of Density Parameters}
\begin{figure}
\centering
\subfloat[Part 1][]{\includegraphics[width=230pt,height=190pt]{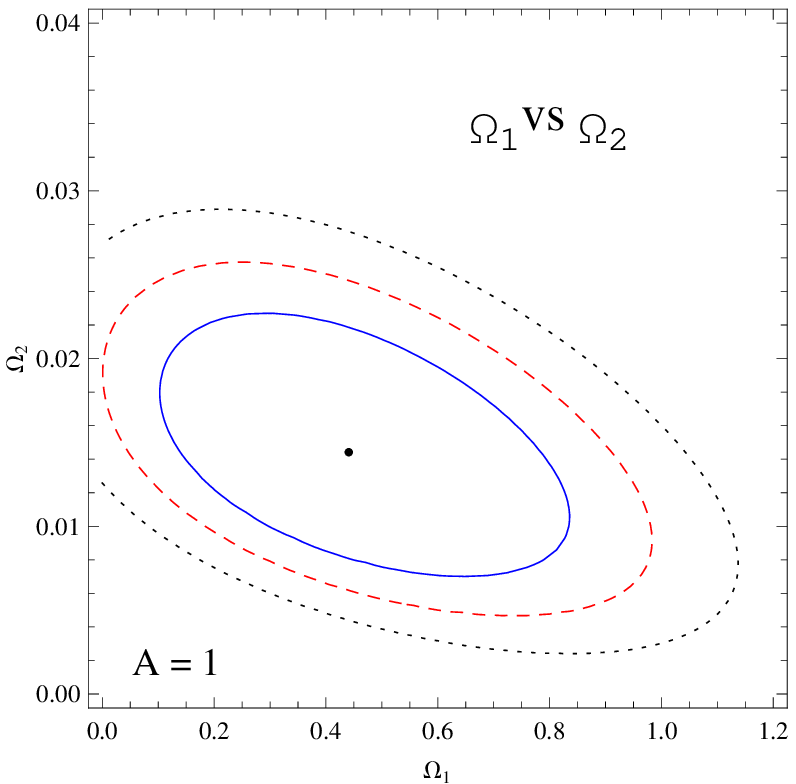} \label{om-a}}\\
\subfloat[Part 2][]{\includegraphics[width=230pt,height=190pt]{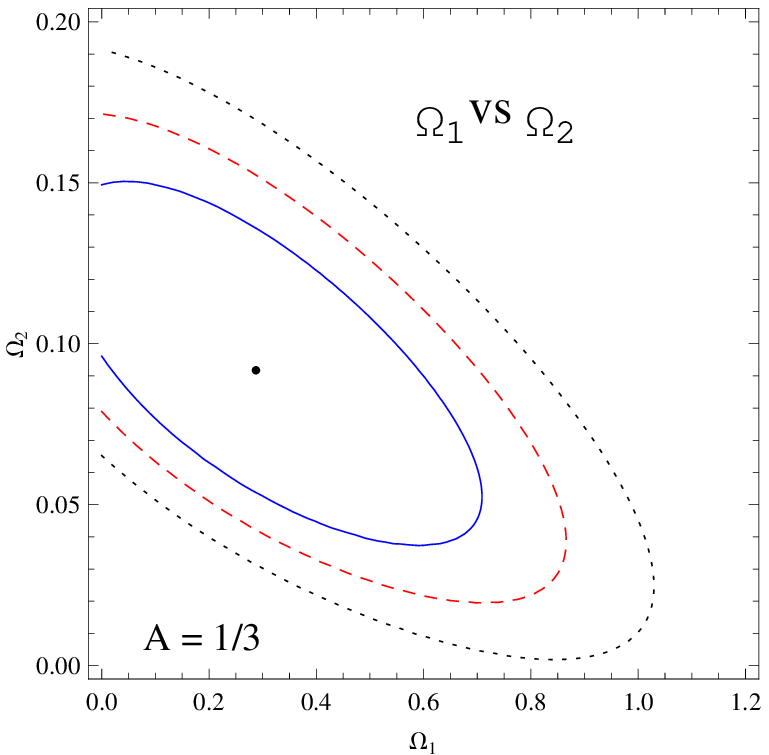} \label{om-b}}\\
\subfloat[Part 3][]{\includegraphics[width=230pt,height=190pt]{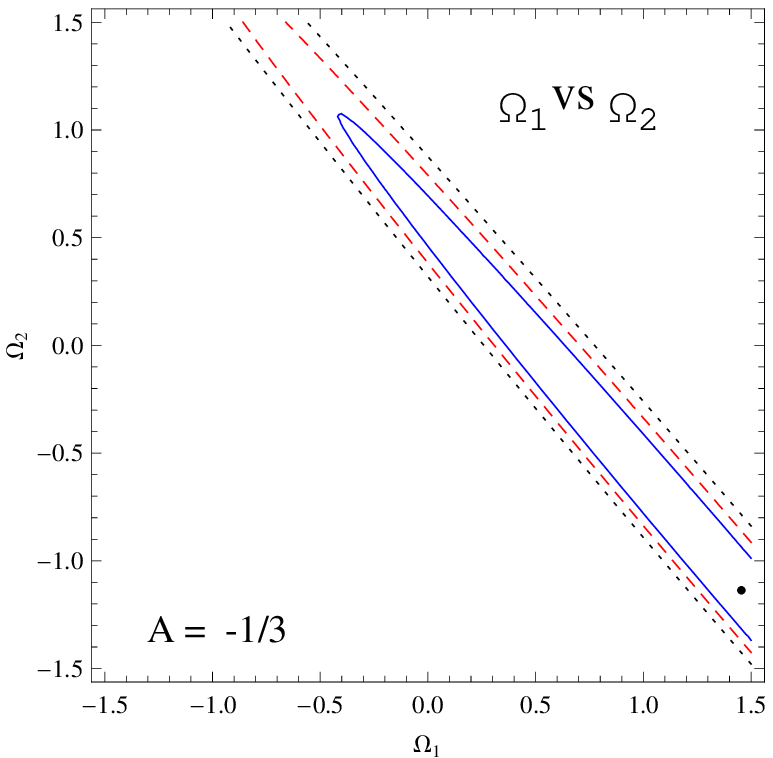} \label{om-c}}\\
\caption[Constraints on DP]{(Colour Online)Contours on $\Omega_{1}-\Omega_{2}$ plane for (a) $A=1$ (b) $A=1/3$ (c) $A=-1/3$:  $66.7\%$(Solid) $95.5\%$ (Dashed) and $99.8 \%$  (Dotted) confidence level.}
\label{omega}
\end{figure}
In the previous section we determined the best fit values for $B$ and $K$ corresponding to different models evoked by different choices of $A$. We now plot contours on $\Omega_{1}-\Omega_{2}$ plane. The $66.7\%$ (solid), $95.5\%$ (dashed) and $99.8\%$ (dotted) contours are shown in fig.(\ref{omega}). EU model with $A=1$ permits a composition of dark energy ($\Omega_{\Lambda}$), dust ($\Omega_1$) and stiff matter ($\Omega_2$) \citep{eu}. For $A=1/3$, $\Omega_1$ represents DP for cosmic strings and $\Omega_2$ represents DP for radiation. For the model with $A=-1/3$, $\Omega_1$ and $\Omega_2$ represent DP for domain wall and cosmic string respectively. The best fit values for $\Omega_{1}$ and $\Omega_{2}$ for different models are obtained which in turn determines  the best fit values for $\Omega_{\Lambda}$  in the corresponding model as:
\begin{equation}
\label{om}
\Omega_{\Lambda}=1-\Omega_{1}-\Omega_{2}
\end{equation}
We have shown the best fit values for the parameters of EU in table(\ref{tomega})
\section{Discussion}

\begin{table}
\centering
		\begin{tabular}{@{}lccc}
		\hline
		Model & $\Omega_1$ & $\Omega_2$ & $\Omega_{\Lambda}$\\
		\hline
		$A=1$ & 0.441 & 0.014 & 0.545\\
		$A=1/3$ & 0.287 & 0.091 & 0.622 \\
		$A=-1/3$ & 1.455 & -1.137 & 0.682\\
		\hline	
		\end{tabular}
	\caption{Findings: Analysis of Density Parameters}
	\label{tomega}
\end{table}
In this work we obtained observational constraints on the model parameters for a class of EU solutions. Here we  consider different values of $A$ belonging to a class of EU given by \citet{eu}. The model parameters
of the EU are constrained using the observed Hubble data (OHD) as well as using a joint analysis with the measurement of a BAO peak parameter. We use BAO peak parameter as suggested by
 \citet{bao} which is independent of dark energy model. The present day value for the density parameters are determined. For the case $A=-1/3$ we have found that the present values of the DP are unrealistic. 
However, in the original work \citet{eu} showed that for the case $A=-1/3$, evolution of the cosmic fluid mimics a composition of dark energy (cosmomological constant), domain walls and cosmic strings. 
As the evaluated value of the present day Density parameter in the model with $A=-1/3$ is not realistic the model with $A=-1/3$ may be ruled out. However, in the other two cases, namely for $A=1$ and
$A=1/3$, we obtain cosmological models with physically realistic  density parameter. The best fit values for the model parameters $B$ and $K$  are determined. It is found that
the model admits dark energy density close to that predicted by observations in $\Lambda$CDM cosmology. The analysis we adopted here involves kinematics only and it would be interesting to analyze and determine the model constraints using the dynamical aspects like structure formation etc. Also it is worthwhile to note that the parameter $K$ should in principle be fixed from the initial conditions itself. A more stringent constraint on the EU may be obtained for a  viable candidate for cosmology. All these issues will be considered elsewhere.

\section*{Acknowledgement}
SG would like to thank CSIR for awarding Senior Research Fellowship. BCP and PT would like to thank IUCAA Resource Centre, NBU for providing research facilities. BCP would like to thank UGC, New Delhi for financial support (Grant No. 36 365/(SR) dated 28 Mar., 2009).

\end{document}